\documentclass[11pt]{article}
\usepackage{epsfig}
\usepackage{amssymb}
\usepackage{amsmath}
\usepackage{graphicx}
\usepackage{latexsym, cite}
\usepackage{graphicx}
\def\draftversion{N}                
\def\note[#1]#2{\message{(#1)}\if\draftversion
{\noindent\em[#2]\/}\fi}
\if \draftversion Y
\fi
\catcode`\@=11
\catcode`\@=12

\newcommand{\eq}[1]{\begin{equation}\label{#1}}
\newcommand{\en}{\end{equation}}
\newcommand{\qe}{\end{equation}}
\newcommand{\ear}[1]{\begin{eqnarray}\label{#1}}
\newcommand{\eqa}[1]{\begin{eqnarray}\label{#1}}
\newcommand{\rae}{\end{eqnarray}}
\newcommand{\ena}{\end{eqnarray}}
\newcommand{\beq}[1]{\begin{equation}\label{#1}}
\newcommand{\eeq}{\end{equation}}
\newcommand{\bea}[1]{\begin{eqnarray}\label{#1}}
\newcommand{\eea}{\end{eqnarray}}

\newcommand{\dfk}{d^4 k}

\newcommand{\exxp}[1]{\exp\left\{ #1 \right\}}
\newcommand{\cutoff}[1]{\frac{#1}{\Lambda^2}}
\newcommand{\pt}{\tilde p}

\def\be{\begin{equation}}
\def\ee{\end{equation}}

\def\b{\beta}
\def\d{\delta}

\def\l{\lambda}
\def\m{\mu}

\def\d{\delta}

\begin{document}
\begin{titlepage}
\vskip0.5cm
\begin{flushright}
AEI-2006-020\\
IMSC-2006-03-7\\
DIAS-STP-06-05 \\
\end{flushright}
\vskip0.5cm
\begin{center}
{\Large\bf  Nonlocal regularisation of noncommutative field theories}
\end{center}
\vskip1.3cm
\centerline{T.~R.~Govindarajan\footnote{On leave of absence from the Institute 
of Mathematical Sciences, CIT Campus, Taramani, Chennai 600113, India; e-mail:
\texttt{trg@imsc.res.in}}$^{a}$, 
Se\c{c}kin~K\"{u}rk\c{c}\"{u}o\v{g}lu$^{b}$ and Marco~Panero$^{b}$}
\vskip1.0cm

\centerline{\sl  $^a$ Max-Planck-Institut f\"ur Gravitationsphysik}

\centerline{\sl Albert-Einstein-Institut}

\centerline{\sl Haus 5, Am M\"uhlenberg}

\centerline{\sl D-14476 Golm, Potsdam, Germany}

\begin{center}
{\sl  e-mail:} \hskip 6mm \texttt{govind@aei.mpg.de}
\end{center}

\vskip0.4 cm

\centerline{\sl  $^b$  School of Theoretical Physics}

\centerline{\sl  Dublin Institute for Advanced Studies}

\centerline{\sl  10 Burlington Road}

\centerline{\sl  Dublin 4, Ireland}
\begin{center}
{\sl  e-mail:} \hskip 6mm \texttt{seckin@stp.dias.ie}{\sl ,} \hskip 2mm \texttt{panero@stp.dias.ie}
\end{center}

 \vskip1.0cm

\begin{abstract}
We study noncommutative field theories, which are inherently nonlocal,
using a Poincar\'e-invariant regularisation scheme which yields an
effective, nonlocal theory for energies below a cut-off scale. After
discussing the general features and the peculiar advantages of this
regularisation scheme for theories defined in noncommutative spaces,
we focus our attention onto the particular case when the
noncommutativity parameter is inversely proportional to the
square of the cut-off, via a dimensionless parameter $\eta$. 
We work out the perturbative corrections at one-loop order for a scalar
theory with quartic interactions, where the signature of
noncommutativity appears in $\eta$-dependent terms. The implications of this approach, which avoids the problems related to UV-IR mixing, are discussed from the perspective of the Wilson renormalisation program. Finally, we remark about the generality of the method, arguing that it may lead to phenomenologically relevant predictions, when applied to realistic field theories.
\end{abstract}
\end{titlepage}

\setcounter{footnote}{0}
\def\thefootnote{\arabic{footnote}}

\section{Introduction}
\label{introsect}

In recent times there has been a growing interest to study quantum field theories in noncommutative spacetime~\cite{Connes,Madore:book}. The major reason is that the underlying spacetime structure when quantum gravity effects are included could be more complex with respect to commutative spacetimes, and may induce 
nonlocal effects~\cite{Doplicher}. The study of NC field theories became 
particularly 
popular after it was realised that they arise 
naturally while considering  open strings in background magnetic fields~\cite{Seiberg:1999vs,CDS}. 

Another class of noncommutative field theories arises in the context of the fuzzy discretisation of certain spaces~\cite{Madore:1991bw,Balachandran:2005ew}. Field theories formulated on fuzzy spaces, such as the fuzzy sphere and the fuzzy ${\mathbb C}P^N$, are essentially finite-dimensional matrix models. Fuzzy spaces preserve the crucial symmetries of the corresponding continuum manifolds: they allow a mathematically well-defined treatment of spinors, gauge fields, as well as topologically non-trivial field configurations~\cite{Balachandran:2005ew,fuzzysusy,topology}, they are not affected by the fermion doubling problem~\cite{baltrg}, and, at least in principle, could provide a potential alternative to lattice regularisations.

The study of NC field theories in a general setting could be very complicated.
Therefore, in the present work, we restrict our attention to QFT in the Groenewold-Moyal spacetimes. 

The expectation that noncommutativity could regularise the divergences of QFT was proven to be false long ago~\cite{Filk:1996dm}: the divergences of a subclass of loop diagrams still survive, and one should regularise and renormalise the theories to fully understand their implications. But at this point a new behaviour, entirely due to noncommutativity, namely the UV-IR mixing, creates an obstruction to the Wilson approach to renormalisation.\footnote{For a discussion on the role of noncommutativity in critical phenomena, see~\cite{Chen:2001qg}.} To interpret this problem, which is due to the fact that a massless pole appears after the UV degrees of freedom are integrated out, would require embedding the QFT in a NC geometry in a broader setting, like string theory.

On the other hand, in our approach we apply the nonlocal regularisation technique~\cite{Evens:1990wf,joglekar} to NC field theory. Pioneering work using the nonlocal regularisation was done in~\cite{Moffat:1990jj,eliezer}; the approach was then applied to QED~\cite{Evens:1990wf}, to supersymmetric models~\cite{Kleppe:1990sa}, and to theories treated with the Batalin-Vilkovisky formalism~\cite{Paris:1996jg}. The philosophy underlying our present program can be described as follows: nonlocality is represented through a form factor with a cut-off 
related to an inherent granular structure of spacetime, which is expected to be due to quantum effects of gravity or to an intrinsic mass scale given by $\Lambda$, then the
resulting model can be interpreted as an effective theory for energies lower than $\Lambda$~\cite{joglekar}\footnote{A different approach to the interplay between NC effects and nonlocal fields was discussed in~\cite{GorbarHashimotoMiransky}.}.
In our approach, the way the cut-off is introduced to regularise the theory induces an intrinsic nonlocality in addition to the noncommutativity. The program allows to avoid the unitarity problems, since it does not induce additional poles in the propagator. It is also worthwhile to note that, in the absence of noncommutativity, this nonlocality is fully compatible with Lorentz invariance, even at a finite cut-off.

In the following, we shall particularly focus on a remarkably interesting case, in which the nonlocal regularisation scale $\Lambda$ and the noncommutativity parameter $\theta$ are related, through a dimensionless parameter $\eta$. When $\Lambda$ is taken to infinity, the noncommutativity parameter vanishes like $\Lambda^{-2}$, while $\eta$ is kept fixed. Despite the fact that this procedure seems to be \emph{ad hoc}, it leads to interesting implications at the phenomenological level and lends itself naturally to the Wilson approach to renormalisation. 

In particular, the dependence of some observables on the $\eta$ parameter can be traced to a remnant of an intrinsic spacetime nonlocality at a small length scale (possibly the Planck length), and, depending on the order of magnitude of $\eta$, may lead to phenomenological effects even at energy scales much lower than the Planck energy.

There exist different contexts in which such a relation between the noncommutativity parameter and the regularisation scale can naturally arise. Normally, in QFT the divergences appear when one considers products of operators at the same spacetime point. However, in principle, the localisation of two operators may lead to new phenomena (like black hole production) altering the geometry of spacetime~\cite{Doplicher}, and noncommutativity is one of the possible effects. Another context where the NC scale could become related to the cut-off occurs for quantum Hall (QH) systems. In the QH setting one assumes that the energy separation among Landau levels is much larger than the scale of the typical interactions between electrons. Therefore, a regularisation procedure for an effective NC theory for the QH system at low energies should allow for the NC scale to be related to the cut-off.

In the light of the recent developments of ``twisted'' Poincar\'e covariance of noncommutative field theories~\cite{Chaichian:2004za,wess}, it would be interesting to develop the above program in a completely covariant fashion: this issue will be addressed in future publications~\cite{newpaper}.

This paper is organised as follows: in section~\ref{formalismsect}, we outline the 
general formalism of nonlocal regularisation, following the works in the literature. In section~\ref{phi44sect}, we consider the $\phi^4$ scalar field theory 
in $D=4$, and present the mass, field strength and coupling renormalisations at one loop order. In section~\ref{betafunctionsect}, we derive the $\beta$-function at the same order, and comment on its behaviour in different regimes --- including the commutative limit, reproducing the standard result. In the final section~\ref{conclusionsect}, we summarise the main results of our work, 
commenting on their physical implications, and on possible future research directions. 
In particular, when the present program is carried out in NC QED, one could draw phenomenological consequences~\cite{newpaper}, which may be compared to the results of present high-energy physics experiments, imposing constraints on the possible values of the NC scale. Some details regarding the calculations in the paper are collected in the Appendix.

\section{General formalism}
\label{formalismsect}

Let us first briefly present the essential features of the nonlocal regularisation of a field theory, following the formalism and notation of~\cite{Kleppe:1991ru,Evens:1990wf}.

We start with an action for a real scalar field $\phi$ in the form:
\eq{scalaraction} 
S[\phi]~=~F[\phi]~+~I[\phi] \; ,
\en
where $F[\phi]$ consists of the sum of the (standard) kinetic and quadratic terms: 
\eq{quadratic}
F[\phi]~=~\frac{1}{2}\int d^D x ~\phi \left( \partial^2 - m^2 \right) \phi \; ,
\en
while $I[\phi]$ is the interaction term. 

Throughout the article, we work in the euclidean signature. 

The nonlocal action can be obtained introducing a finite momentum scale $\Lambda$  to define the ``smeared'' field $\hat{\phi}$ as: 
\eq{phihatdefinition}
\hat{\phi}~=~\exp \left(-\frac{p^2+m^2}{2\Lambda^2} \right)~\phi \; .
\en
Then the action can be rewritten as a function of the smeared field $\hat{\phi}$ and of a ``shadow'' field $\psi$:
\eq{nonlocalaction}
S[\phi,\psi]~=~F[\hat{\phi}]~-~A[\psi]~+~I[\phi~+~\psi]
\ee
where:
\eq{auxiliaryfieldaction}
A[\psi]=~\frac{1}{2} \int d^D x~\psi~\left[\frac{\cal F}{{\cal{E}}^2~-~1} \right]~\psi \, ,
\en
with: ${\cal E}=\exp\left( \frac{{\cal{F}}}{2 \Lambda^2} \right)$ and: ${\cal{F}}=-(p^2+m^2)$. The theory can be regularised after eliminating $\psi$ via its equations of motion:
\eq{snonlocal}
S_{\mbox{\tiny{nl}}}[ \phi]~=~S[\phi,\psi]|_{\psi=\psi_{\mbox{\tiny{cl}}}(\phi)} \;\;\; , \mbox{ with:} \;\;\; \left. {\frac{\d S}{\d\psi}}\right|_{\psi=\psi_{\mbox{\tiny{cl}}}}=0 \;.
\ee
Essentially, the procedure amounts to separate the original $\phi$ propagator associated with any internal line into the sum of a ``smeared'' propagator: 
\eq{smearedpropagator}
-\frac{1}{p^2+m^2}\exp\left(-\frac{p^2+m^2} {\Lambda^2}\right) \; , 
\en
and the propagator for the shadow field $\psi$:
\eq{shadowfieldpropagator}
-\frac{1}{p^2+m^2}\left[ 1 - \exp\left(-\frac{p^2+m^2}
{\Lambda^2}\right) \right] \; .
\en
The Feynman rules are changed accordingly: each line in the internal loops can be either the propagator of a smeared or of a shadow field; the only restriction is that internal loops made solely out of shadow field propagators are not allowed.
  
Let us now consider a scalar field theory defined on the $D$-dimensional Groenewold-Moyal spacetime. First, we recall that
the latter is defined to be  the algebra of functions ${\cal A}_\theta({\mathbb R}^D)$ generated by the coordinates 
$x^\mu \, (\mu \in \{ 0,1,2, \cdots , D-1 \} )$, with the commutation relation:
\eq{nc}
\left[x^\mu, x^\nu \right]_* = i \theta^{\mu\nu} \;,
\en
with $*$ denoting the Groenewold-Moyal star-product. For the functions $f\,, g \in {\cal A}_\theta({\mathbb R}^D)$ this $*$-product is defined as~\cite{Szabo:2001kg}:
\eq{starproduct}
f*g(x)~=~f(x) e^{\frac{i}{2} \overleftarrow{\partial}_\mu \theta^{\mu\nu}\overrightarrow{\partial}_\nu}g(x) \; .
\en
We note that as the theory for a free scalar field in NC theory is the same as that of the commutative theory, the above general discussion goes through for the NC theory as well without any changes.

We restrict our attention to interactions which are polynomials in the $*$-product of $\phi$. A typical interaction term can be expressed as: 
\eq{interactionterm}
I[\phi]~ = \lambda \int d^D x~ \phi * \phi * \cdots * \phi  \;.
\en

We note that the standard phase factor~\cite{Minwalla:1999px}:
\eq{phasefactor}
\exp \left( -\frac{i}{2} \sum_{i<j} p^i_\mu \theta^{\mu\nu} p^j_\nu \right) \; ,
\en 
that appears in the interaction vertex of $\phi * \phi * \cdots * \phi$, contributes 
to all diagrams, regardless the fact that the internal loops carry smeared or shadow field 
propagators.

Starting from the next section, we shall focus our attention only to quartic interactions of this form in four spacetime dimensions.

\section{Noncommutative $\phi^4$ theory}
\label{phi44sect}

Here we focus our attention to the theory with quartic interactions:
\eq{ncquarticinteraction}
I(\phi)~=~\frac{\l}{4!}~\int~d^4 x ~\phi * \phi * \phi *\phi \;.
\en 
This model allows us to capture all the essential features coming from our combined treatment of noncommutativity and nonlocality, along the lines presented above.

We write $\theta^{\mu\nu}$ in the form:
\eq{thetadefinition}
\theta^{\mu\nu} = \theta
\left (
\begin{array}{ccc} 
0 & 1 & \, \\
-1 & 0 & \, \\
\, & \, & \ddots 
\end{array}
\right) \, 
\en
and, in what follows, we assume that $\theta$ is inversely proportional to $\Lambda^2$, via a dimensionless parameter $\eta$:
\eq{etadefinition}
\theta = \frac{\eta}{\Lambda^2} \; .
\en
As we shall see later on, this assumption will allow us to probe a limit of the theory, where UV-IR mixing is no longer present.

In the following subsections, we address the one-loop perturbative calculation of the 
propagator and vertex renormalisation, using the Feynman rules reviewed in section~\ref{formalismsect}, and according 
to the assumption in eq.~(\ref{etadefinition}).

\subsection{Propagator renormalisation} 
\label{propagatorsubsect} 

\begin{figure}[t]
\setlength{\unitlength}{1cm}
\begin{minipage}[t]{7.5cm}
\begin{center}
\begin{picture}(6.0,5.0)
\epsfig{file=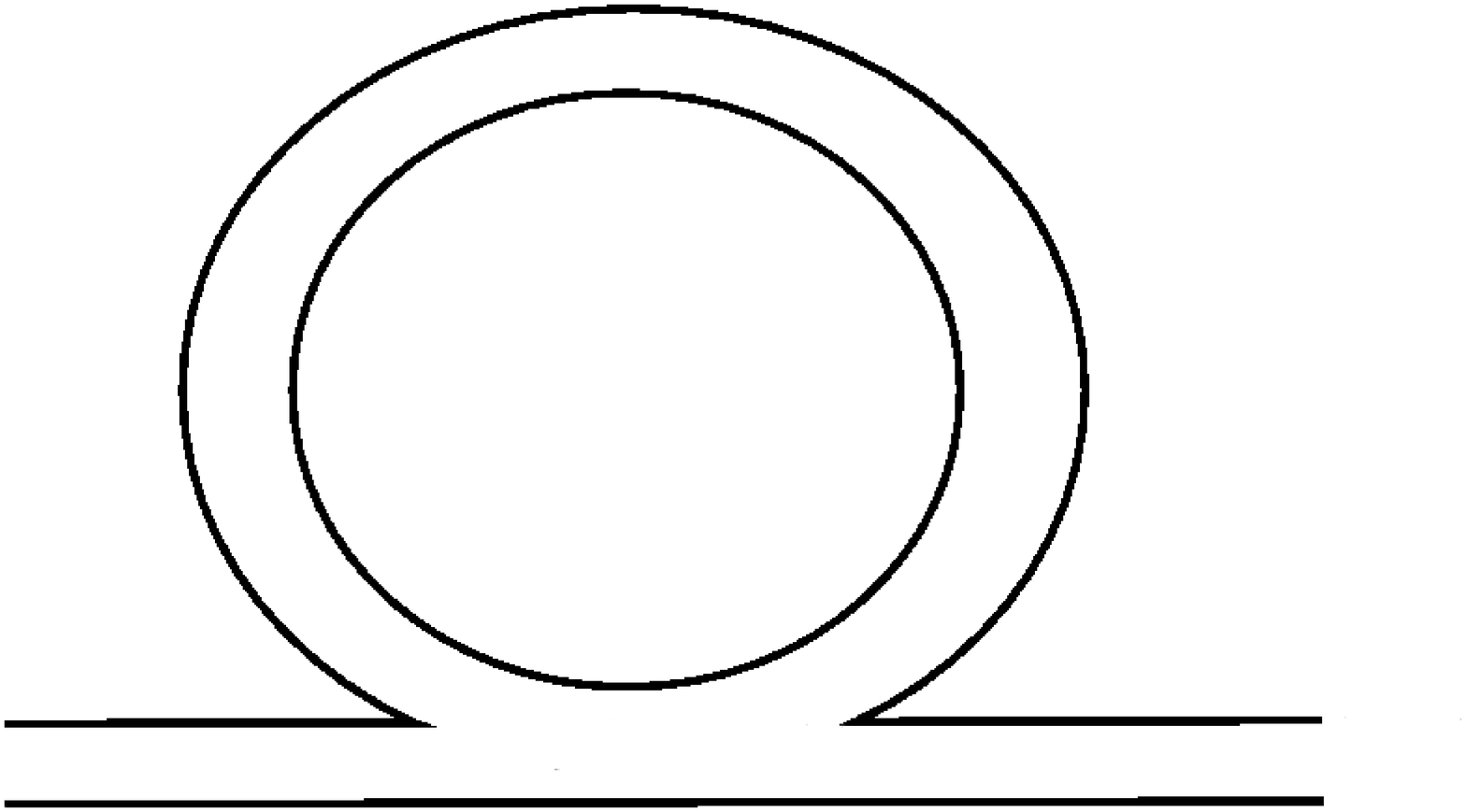,width=\hsize}
\end{picture}\par
\end{center}
\caption{\label{tadpolefig} The planar tadpole diagram.}
\end{minipage}
\hfill
\begin{minipage}[t]{7.5cm}
\begin{center}
\begin{picture}(6.0,5.0)
\epsfig{file=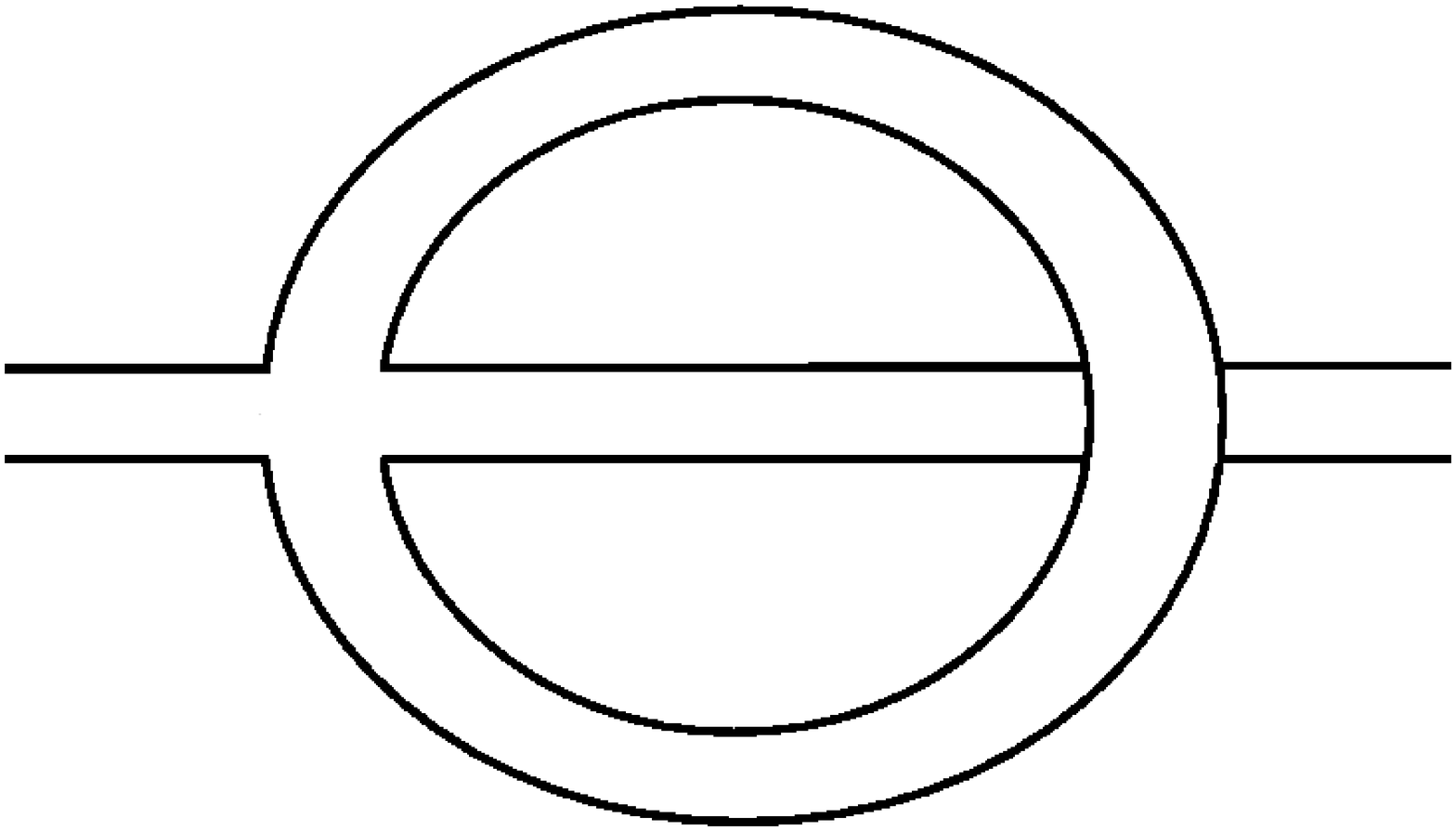,width=\hsize}
\end{picture}\par
\end{center}
\caption{\label{nonplanartadpolefig} The non-planar version of the tadpole diagram.}
\end{minipage}
\end{figure}

First, we consider the renormalisation of the propagator, up to one loop. 
The Feynman diagrams contributing to the one-loop self-energy are depicted in double-line notation in fig.~\ref{tadpolefig} and in fig.~\ref{nonplanartadpolefig}.
For these diagrams, the internal loop is necessarily associated with a smeared propagator, since diagrams containing loops built from shadow propagators only are not allowed in the theory. 

The contributions of the two diagrams in fig.~\ref{tadpolefig} and in fig.~\ref{nonplanartadpolefig}, including their multiplicity factors are, respectively:
\eq{tadpoleintegral}
I_{\mbox{\tiny{planar}}}~=~\frac{\l}{3 (2 \pi)^4}\int d^4k~\frac{e^{-\frac{k^2+m^2}{\Lambda^2}}}{k^2+m^2} \,,
\en
and
\eq{nonplanartadpoleint}
I_{\mbox{\tiny{non-planar}}}~=~\frac{\l}{6 (2 \pi)^4}\int d^4k~\frac{e^{-\frac{k^2+m^2}{\Lambda^2}}}{k^2+m^2}e^{ik_{\mu}\theta^{\mu\nu}p_\nu} \,.
\ee
When $\Lambda \rightarrow \infty$, eq.~(\ref{tadpoleintegral}) is quadratically divergent, while eq.~(\ref{nonplanartadpoleint}) 
converges by virtue of the oscillating factor, provided that neither $\theta$ nor {the external momenta are zero}.
 
These integrals can be evaluated in a straightforward way. Taking into account eq.~(\ref{etadefinition}), eq.~(\ref{tadpoleintegral}) and eq.~(\ref{nonplanartadpoleint}) yield:
\begin{equation}
\Sigma(p^2) = \frac{\lambda}{2^5 \pi^2}
\left\lbrace \Lambda^2 - m^2 \ln \left(\cutoff{m^2} \right) + m^2 (\gamma -1) 
-\frac{1}{6} \eta^2 |\pt|^2 \right \rbrace + O \left(\cutoff{1} \right) \,,
\label{Itotal}
\end{equation}  
having defined: ${\tilde p}^\nu = \frac{1}{2\theta} p_\mu \theta^{\mu \nu}$. Some details of the calculation leading to eq.~(\ref{Itotal}) are outlined in the Appendix. 

From eq.~(\ref{Itotal}) it is possible to observe that the UV-IR mixing is no longer present, since it is now possible to set $p$ equal to zero in eq.~(\ref{Itotal}) without introducing an infrared singularity. The UV-IR mixing returns only in the limit when $\eta$ and $\Lambda$ are taken to infinity and $\theta=\frac{\eta}{\Lambda^2}$ is kept finite. We postpone the detailed discussion of this and various other limiting cases until section~\ref{conclusionsect}.

For the sake of completeness and to fix the notations and conventions, we state some common definitions on generalities of renormalisation. The field renormalisation constant $Z$ and the bare parameters $m_0$ and $\Lambda_0$ are related to the renormalised mass $m$ and coupling $\lambda$ by: 
\begin{gather}
Z = 1 + \delta Z(\lambda, m^2, \Lambda^2) \,, \\
Z m_0^2 = Z m^2 + \delta m^2 (\lambda, m^2, \Lambda^2) \,, \\
Z^2 \lambda_0 = \lambda + \delta \lambda (\lambda, m^2, \Lambda^2) \,.
\end{gather}
Denoting the renormalised self-energy (1PI two-point function) by $\Sigma_R(p^2)$, the mass and field renormalisations are determined by:
\be
\Sigma_R (-m^2) = 0 \,, \quad \left.{\frac{\partial \Sigma_R(p^2)}{\partial p^2} }\right|_{p^2 = - m^2} = 0 \,.
\ee
The renormalised coupling is determined by the value of the renormalised 1PI four-point function $V_R(p_1, p_2, p_3, p_4)$ for a particular choice of external momenta. In the next subsection we shall chose:
\begin{equation}
\left.{V_R}\right|_{p_1 = p_2 = p_3 = p_4 = q } = - \lambda \,.
\label{VR}
\end{equation}

Using these renormalisation conditions and the contribution to the self-energy $\Sigma(p^2)$ given by eq.~(\ref{Itotal}), it is straightforward to extract the mass and field strength renormalisations:\footnote{In anticipation of the fact that $\delta Z = O(\lambda^2)$, the powers of $Z$ appearing in the loop integrals and elsewhere are suppressed, since they are all equal to $1$
at the one loop order of the perturbation theory.}
\begin{eqnarray}
\delta m^2 &=& \frac{\lambda}{2^5 \pi^2}
\left\lbrace - \Lambda^2 + m^2 \ln \left(\cutoff{m^2} \right) + m^2 (1 - \gamma)  
+ \frac{1}{6} \eta^2 |\pt|^2 \right \rbrace + O \left(\cutoff{1} \right) \,, \nonumber \\
\delta Z &=& O(\lambda^2) \,.
\label{massrenormalisation}
\end{eqnarray}
From eq.~(\ref{massrenormalisation}) it is readily observed that the mass renormalisation is essentially the same as that of the commutative $\phi^4$ theory, except that now its finite part receives a correction depending on $\eta$, as a remnant of the noncommutativity of the theory.

\subsection{Vertex renormalisation}
\label{vertexsubsect} 

\begin{figure}[t]
\setlength{\unitlength}{1cm}
\begin{minipage}[t]{7.5cm}
\begin{center}
\begin{picture}(6.0,5.0)
\epsfig{file=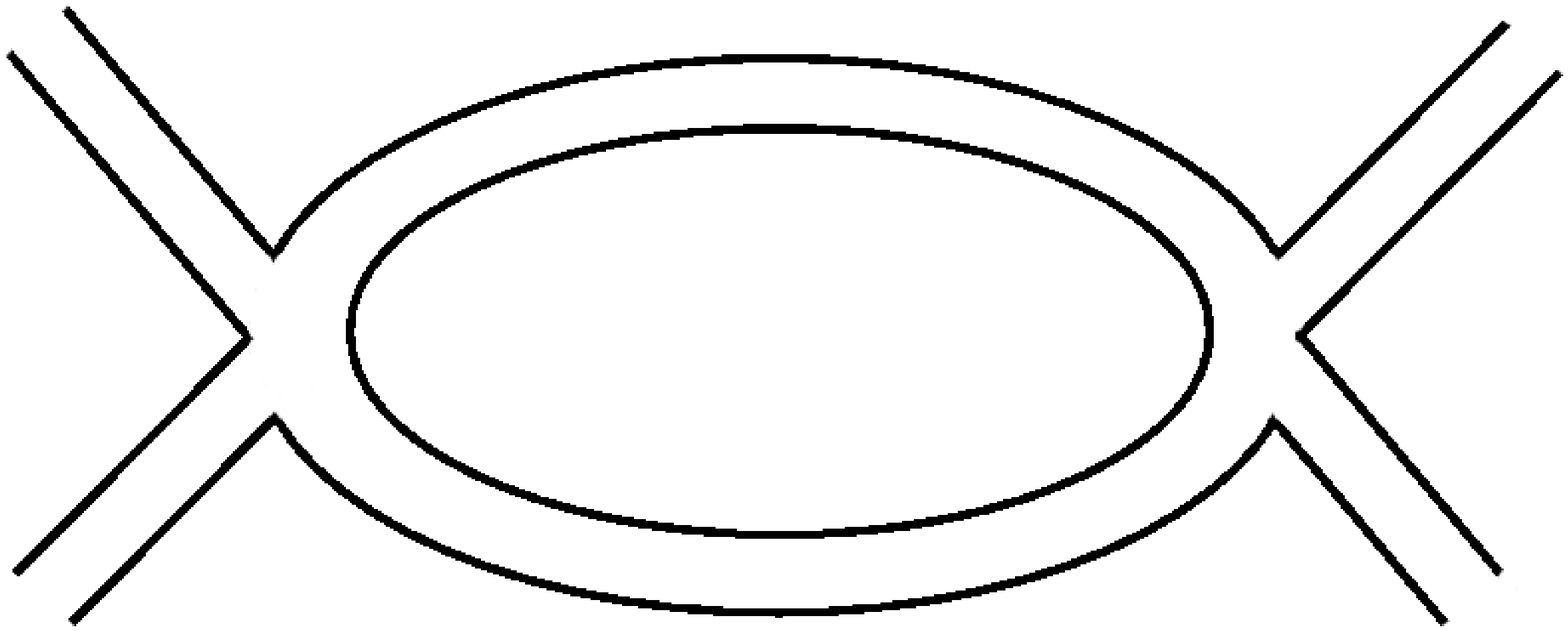,width=\hsize}
\end{picture}\par
\end{center}
\caption{\label{fishfig} The ``fish'' diagram.}
\end{minipage}
\hfill
\begin{minipage}[t]{7.5cm}
\begin{center}
\begin{picture}(6.0,5.0)
\epsfig{file=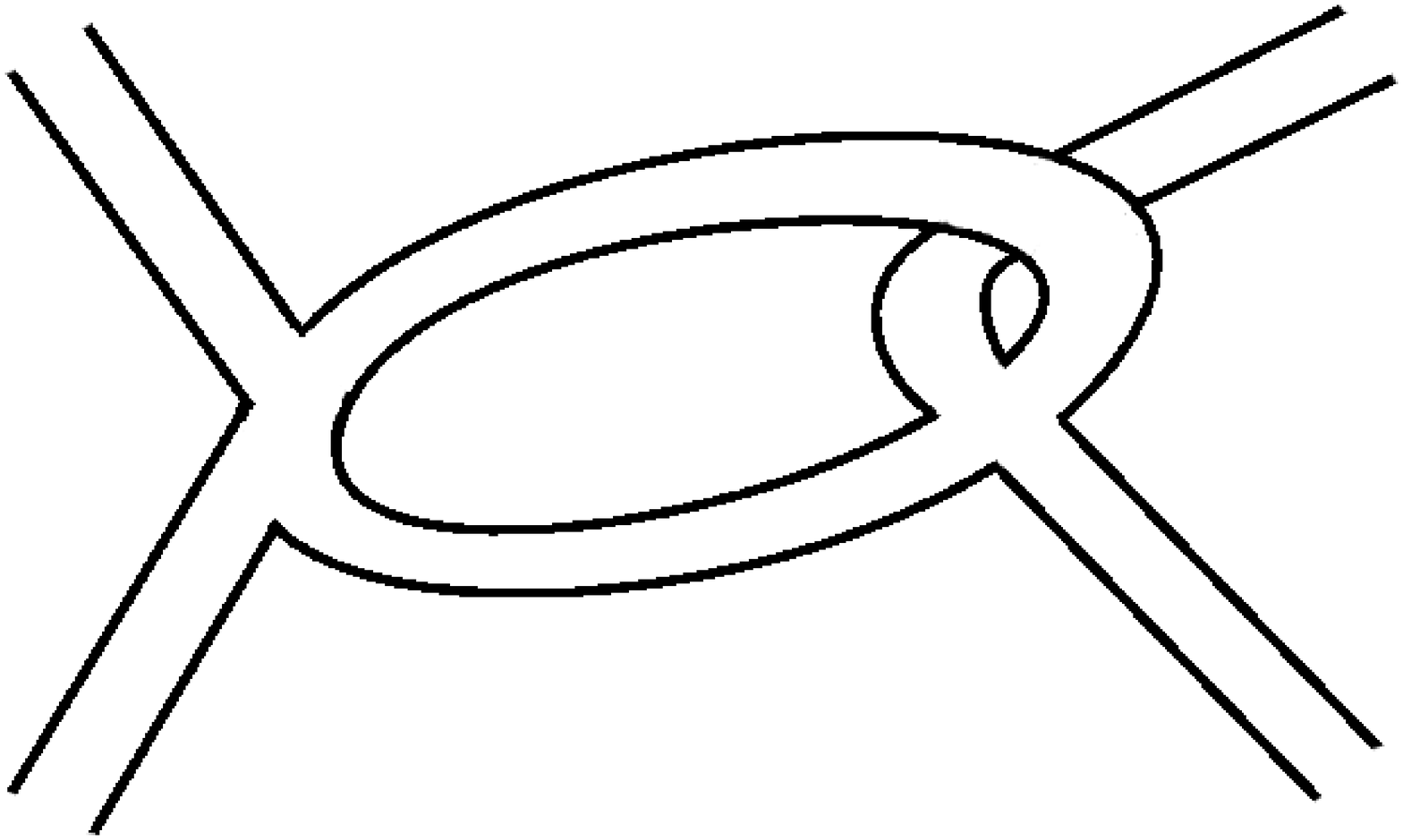,width=\hsize}
\end{picture}\par
\end{center}
\caption{\label{talkingfishfig} The ``talking fish'' diagram.}
\end{minipage}
\end{figure}

\begin{figure}[t]
\setlength{\unitlength}{1cm}
\begin{minipage}[t]{7.5cm}
\begin{center}
\begin{picture}(6.0,5.0)
\epsfig{file=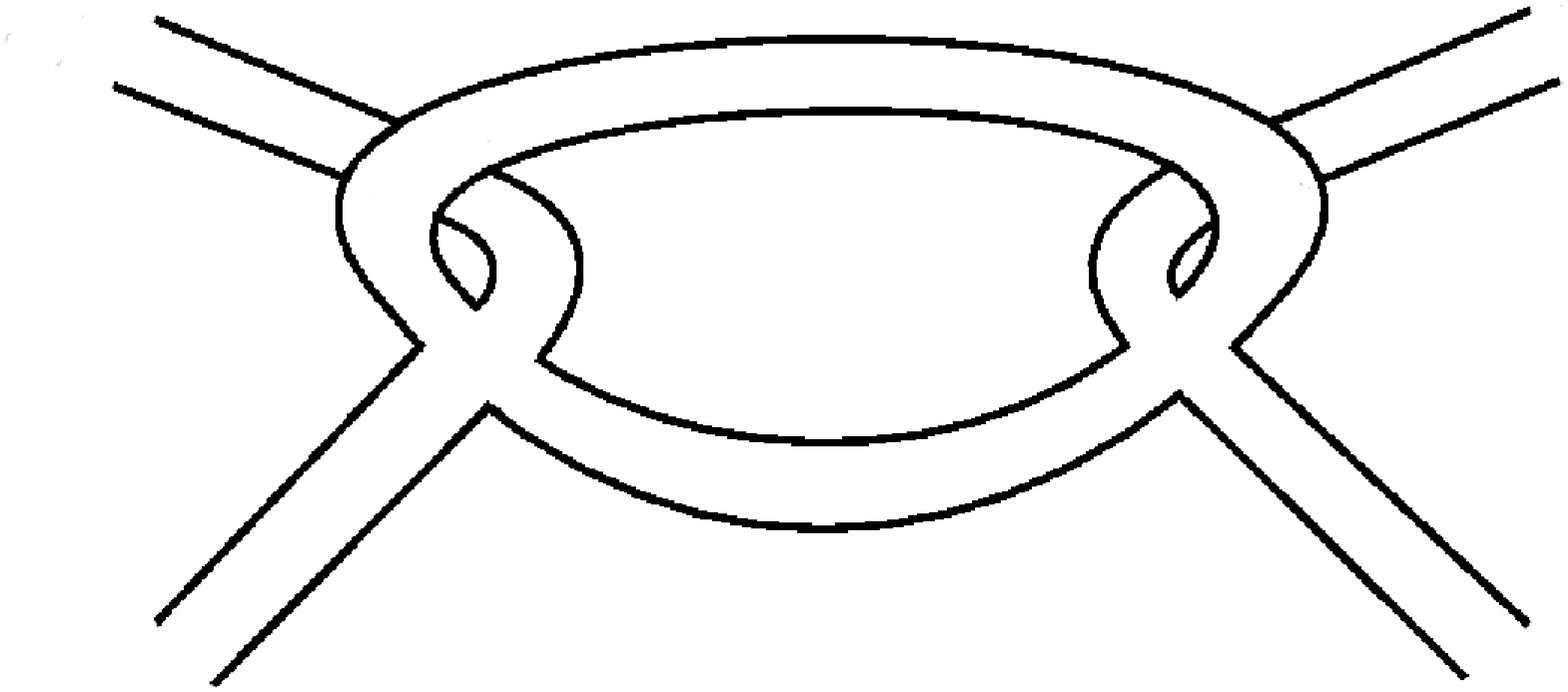,width=\hsize}
\end{picture}\par
\end{center}
\caption{\label{linkfig} The ``link'' diagram.}
\end{minipage}
\hfill
\begin{minipage}[t]{7.5cm}
\begin{center}
\begin{picture}(6.0,5.0)
\epsfig{file=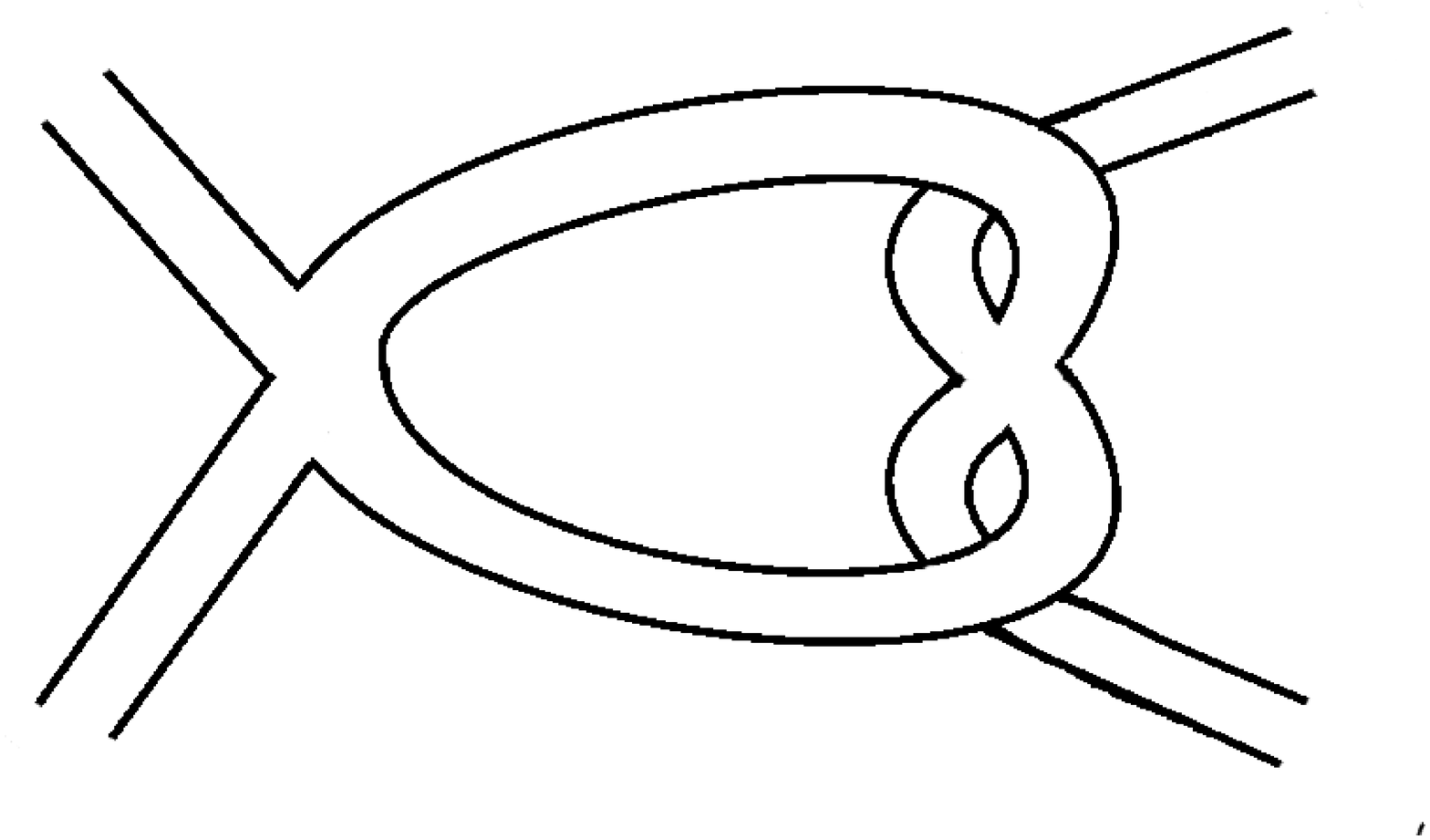,width=\hsize}
\end{picture}\par
\end{center}
\caption{\label{heartfig} The ``heart'' diagram.}
\end{minipage}
\end{figure}

We now consider the renormalisation of the coupling constant up to one-loop order. In addition to the four-point vertex, the
relevant diagrams contributing to the 1PI four-point function are shown in fig.~\ref{fishfig}, \ref{talkingfishfig}, \ref{linkfig} and \ref{heartfig}. 
In all of these diagrams, the internal lines could both carry the smeared field propagator, or it could be that one or the other is associated with the shadow propagator.\footnote{Note that there are two equivalent contributions from the latter, and  the total contribution of each diagram is evaluated by adding all of the three.}

The typical loop integrals associated to non-planar diagrams are:
\eq{i2}
\frac{\lambda^2}{(4!)^2} \frac{1}{(2 \pi)^4} \int \frac{\dfk}{(k^2 +m^2)[(k-P)^2+m^2]} \exxp{-\cutoff{1} \left[ k^2+ m^2 + (k-P)^2+m^2 \right] + i Q_\mu \theta^{\mu\nu} k_\nu} \,,
\ee
\eq{i3}
\frac{\lambda^2}{(4!)^2} \frac{1}{(2 \pi)^4} \int \frac{\dfk \left\{ 1 - \exp \left[ -\cutoff{1} \left( k^2+ m^2 \right) \right] \right\}} {(k^2 +m^2)[(k-P)^2+m^2]} \exxp{-\cutoff{1} \left[ (k-P)^2+m^2 \right] + i Q_\mu \theta^{\mu\nu} k_\nu} \, ,
\ee
where $P=p_1+p_2$ is the sum of the incoming momenta, and $Q$ denotes one of the momenta attached to the external legs, or a sum thereof.\footnote{We denote the incoming momentum attached to the leg in the upper-left corner of the diagrams as $p_1$, and label the incoming momenta associated with the remaining external legs, ordered counter-clockwise, as $p_2$, $p_3$ and $p_4$.}

For the ``fish'' diagram and an unspecified non-planar diagram the results of the loop integrals give:
\begin{gather}
I_{\mbox{\tiny{fish}}} = \frac{\lambda^2}{(4!)^2} \frac{1}{2^4 \pi^2} \left \lbrace \ln \left (\frac{\Lambda^2}{m^2} \right ) + (\ln2 - \gamma - 1) + (2 \ln 2) \left (\frac{m^2}{\Lambda^2} \right ) \right \rbrace 
+ O \left [ \frac{1}{\Lambda^2} \ln \left (\frac{\Lambda^2}{m^2} \right ) \right ]\,, \\
I_{\mbox{\tiny{non-planar}}} = \frac{\lambda^2}{(4!)^2} \frac{1}{2^4 \pi^2} \left \lbrace \ln \left (\frac{\Lambda^2}{m^2} \right ) + (\ln2 - \gamma - 1) + (2 \ln 2) \left(\frac{m^2}{\Lambda^2} \right ) - \frac{3 \eta^2{\tilde Q}^2}{4 \Lambda^2} \right \rbrace
+ O \left [ \frac{1}{\Lambda^2} \ln \left (\frac{\Lambda^2}{m^2} \right ) \right ] \,.
\label{eq:tfloop}
\end{gather}
The results of the ``talking fish'', ``link'' and ``heart'' diagrams are specified by setting ${\tilde Q} = {\tilde p}_i \,, \, (i= 1,2,3,4)$, 
${\tilde Q} = {\tilde p}_2 + {\tilde p}_3$, ${\tilde Q} = ({\tilde p}_2 + {\tilde p}_4)$, or ${\tilde Q} = ({\tilde p}_3 + {\tilde p}_4)$, respectively. Some details of the calculations leading to eq.~(\ref{eq:tfloop}) are outlined in the Appendix.

Summing all these loop integrals with their respective overall multiplicities $m_i~(i=1,2,3,4)$, we find the one loop contribution to the 1PI four-point function:
\begin{equation}
I^T = m_1 \, I_{\mbox{\tiny{fish}}} + m_2 \, I_{\mbox{\tiny{talking fish}}} + m_3 \, I_{\mbox{\tiny{link}}} + m_4 I_{\mbox{\tiny{heart}}} \,.
\end{equation}

The renormalised 1PI four point-function reads:
\begin{equation}
V_R(p_1, p_2, p_3, p_4) = - \lambda + I^T - \delta \lambda \,.     
\label{VR1}
\end{equation}
Using eq.~(\ref{VR}) in eq.~(\ref{VR1}) and the multiplicities: $m_1~=~4!8$, $m_2~=~4!16$, $m_3~=4!4$ and $m_4~=~4!8$, we find the renormalisation in the coupling to be:
\eq{couplingrenormalisation}
\delta \lambda = \frac{\lambda^2}{3 \cdot 2^5 \pi^2} \Bigg \lbrace 9 \ln \left ( \frac{\Lambda^2}{m^2} \right ) + 
(18 \ln2) \left (\frac{m^2}{\Lambda^2} \right) + 9 ( \ln2 -\gamma-1) - \frac{12 \eta^2 {\tilde q}^2}{\Lambda^2} \Bigg \rbrace 
+ O \left [ \frac{1}{\Lambda^2} \ln \left (\frac{\Lambda^2}{m^2} \right ) \right ]  \,.
\en

We note that, in the formul\ae~above, the $O \left ( \frac{1}{\Lambda^2} \right)$ terms have been deliberately kept, in order to reveal the dependence of the coupling renormalisation on $\eta$ and the external momenta, which only appear at this order, as it can be easily observed from eq.~(\ref{couplingrenormalisation}). Naturally, this dependence is also present in the $\beta$-function of the theory. We demonstrate this next. 

\section{$\b$-function for the $ \phi^4_*$ theory}
\label{betafunctionsect}

We use the definition of the $\beta$-function employed in~\cite{Kleppe:1991ru}:
\begin{equation}
\beta(\lambda) = m \left (\frac{\partial \lambda}{\partial m} \right)_{\Lambda, \lambda_0}
= - \Lambda \left( \frac{\partial \lambda}{\partial \Lambda} \right)_{m, \lambda_0} \,. 
\end{equation}
Since the derivative of $\lambda_0$ at fixed $\lambda_0$ is zero, we have:
\begin{eqnarray}
\beta(\lambda) &=& \left ( {\frac{\partial \lambda_0}{\partial \lambda}}_{m, \Lambda} \right)^{-1} \Lambda \frac{\partial
  \lambda_0} {\partial \Lambda}_{m, \Lambda} \nonumber \\
  &=& \frac{\lambda^2}{3 \cdot 2^5 \pi^2} \left [ 18 - (36 \ln2) 
  \left(\frac{m^2}{\Lambda^2} \right) + 24 \frac{\eta^2 \tilde q^2}{\Lambda^2} \right] + O \left [ \frac{1}{\Lambda^2} \ln \left (\frac{\Lambda^2}{m^2} \right ) \right ]\,,
\end{eqnarray}
where in obtaining the second line we have made use of eq.~(\ref{couplingrenormalisation}).  

From this result we see that the first term is the same as the $\beta$-function of the commutative $\phi^4$ theory at one loop. The dependence on $\eta$ and on external momenta appears only at the $\frac{1}{\Lambda^2}$ order. However, it now seems possible to choose $\frac{\eta^2 \tilde q^2}{\Lambda^2}$ such that it is comparable to the first term in $\beta(\lambda)$: this indicates a change in the phase structure as momenta become large enough, so that $\eta^2 \tilde q^2$ becomes comparable to $\Lambda^2$. But, viewed as an effective theory for momenta compared to the cut-off $\Lambda^2$, it is expected that new features will arise. We shall comment on this in  section~\ref{conclusionsect}.

\section{Conclusions and perspectives}
\label{conclusionsect}

In this paper we have applied the nonlocal regularisation technique to study NC QFT. The important feature in our approach is the fact that the noncommutativity parameter is related --- through a dimensionless proportionality constant --- to the inverse of the square of $\Lambda$, therefore NC effects (possibly at much lower energies than the Planck mass) arise as due to an intrinsic, very large but finite, energy scale.

Although the idea of addressing a NC theory with a nonlocal regularisation is completely general, and does not rely on this particular assumption, we found that the latter actually proves helpful in eluding the typical problems related to UV-IR mixing, and achieving the Wilson renormalisation program.

The effect of noncommutativity is captured by the $\eta$ parameter, whose phenomenological implications can be successfully worked out. The literature documents the considerable efforts which have been spent, trying to extract phenomenological information from NC theories (see, for example,~\cite{Hinchliffe:2002km}), but most of the standard approaches are spoilt by the UV-IR mixing problem, and hence cannot properly take into account loop effects.

In our treatment, which insists on the Wilson approach to renormalisation, new physics effects become manifest when $\eta^2 {\tilde q}^2$ is comparable to $\Lambda^2$; a similar behaviour is encountered even in theories without noncommutativity, through causality violation~\cite{joglekar}. Such new physics effects in NC spaces are indeed possible, as pointed out in~\cite{Gubser:2000cd}.

The philosophy underlying our program bears some analogies with what is done in fuzzy physics, where the matrix size and the noncommutativity parameter need to be scaled in a proper way, in order to reproduce the continuum limit of a field theory~\cite{Vaidya:2003ew}. 

In the light of the recent developments about the issue of Poincar\'e covariance in NC theories~\cite{wess}, it would be interesting to discuss the problem in the framework of the regularisation presented here, as the latter is {\it per se} covariant. Furthermore, this approach, which allows to deal with a (modified, nonlocal) gauge invariance, also opens up the possibility for a study of QED in a NC geometry, which will be presented elsewhere~\cite{newpaper}.

\vskip1.0cm {\bf \noindent Acknowledgements}\\
\noindent 
We thank S.~Bal, W.~Bietenholz, A.H.~Fathollahi, S.D.~Joglekar, O.~Lisovyy, D.~O'Connor and C.~Saemann for useful discussions. This work was started during a period when T.R.G. was on leave from the IMSc, visiting DIAS; T.R.G.  thanks D.~O'Connor and DIAS for the visit and support. S.K. acknowledges support from the IRCSET postdoctoral fellowship. M.P. acknowledges support received from Enterprise Ireland under the Basic Research Programme.

\newpage
\appendix{}
\label{app}

\section{Appendix}
\label{phi44appsect}

Here we collect some calculation details of the integrals encountered in the text.
We suppress all the prefactors of the integrals to display them in their generality.

\renewcommand{\theequation}{A.\arabic{equation}}
\setcounter{equation}{0}

The integrals contributing to the one-loop two-point function are given in eq.~(\ref{tadpoleintegral}) and eq.~(\ref{nonplanartadpoleint}): they can be easily computed using the Schwinger parametrisation. Denoting the integral appearing in eq.~(\ref{nonplanartadpoleint}) by $I_1$, one gets:
\eqa{nplpropschwinger}
I_1 &=& \cutoff{1} \int_1^\infty d \alpha \int \dfk \exxp{-\cutoff{1} \left[ k^2 +m ^2 -2i \eta \pt k \right]} \nonumber \\
&=& \pi^2 \Lambda^2 \int_1^\infty \frac{d \alpha}{\alpha^2} \exp \left( - \cutoff{m^2} \alpha \right) \exp \left(-\frac{1}{\alpha} \cutoff{\eta^2 \pt^2 } \right) \nonumber \\
&=& \pi^2 \m^2 \sum_{n=0}^\infty \frac{(-1)^n}{n!} \left(\frac{m^2 \eta^2 {\tilde p}^2}{\Lambda^2} \right)^n \Gamma \left(-(n+1) \,, 
\frac{m^2}{\Lambda^2} \right) \nonumber \\
&=& \pi^2 \Lambda^2 + \pi^2 m^2 \ln \left( \cutoff{m^2} \right) +\pi^2 \left[ m^2 \left( \gamma -1 \right) - \frac{\eta^2 \pt^2}{2} \right] + O\left( \cutoff{1} \right) \;. 
\ena
The planar tadpole contribution, apart from the prefactors, can be immediately obtained from eq.~(\ref{nplpropschwinger}), by setting ${\tilde p}$ to zero. 

In the formul\ae~above and in what follows, $\Gamma(m, z)$ denotes the incomplete $\Gamma$-function. We use the recursion relation:
\eq{Gamma} 
\Gamma(n, z) = - \frac{1}{n} z^n e^{-z} + \frac{1}{n} \Gamma (n+1, z)\,,
\ee
and the asymptotic expansion:
\eq{asymptoicexp}
\Gamma(0, z) = -\ln z - \gamma - \sum_{n=1}^\infty \frac{(-z)^n}{n\cdot n!} \,,
\ee
to deduce the value of integrals order by order in $\Lambda$.

Typical integrals associated to the non-planar diagrams of fig.~\ref{talkingfishfig}, \ref{linkfig}, and~\ref{heartfig} are given in eq.~(\ref{i2}) and eq.~(\ref{i3}); we denote the integrals in these expression by $I_2$ and $I_3$, respectively. Using the Schwinger parametrisation, $I_2$ and $I_3$ can be cast into the form:
\eq{kissingfish4}
I_2 \Big|_{P=0} = 2 \pi^2 \int_0^\frac{1}{2} dx \sum_{n=0}^{\infty} \frac{(-1)^n}{n!} \left( \frac{\eta^2 \tilde Q^2 m^2}
{\Lambda^4} \right)^n ~\Gamma \left( -n, \cutoff{m^2} \frac{1}{x}\right) \;,
\ee
\eq{kissingfishshadow4}
I_3 \Big|_{P=0} = \pi^2 \int_\frac{1}{2}^1 dx \int_{\cutoff{m^2} \frac{1}{x}}^{\cutoff{m^2} \frac{1}{1-x}} \frac{dt}{t} e^{-t} \sum_{n=0}^{\infty} \frac{(-1)^n}{n!} \left(\frac{\eta^2 \tilde Q^2 m^2}{\Lambda^4}\frac{1}{t} \right)^n \,.
\ee
We use eq.~(\ref{Gamma}) and eq.~(\ref{asymptoicexp}) to deduce the value of these integrals order by order in $\Lambda$.

\end{document}